\documentclass[preprint,prb]{revtex4}
\usepackage{amsfonts}
\usepackage{amsmath}
\usepackage{amssymb}

\begin{document}

\title
{REVISITING THE FR\"{O}HLICH-TYPE TRANSFORMATION WHEN DEGENERATE STATES ARE PRESENT.
}%

\author{M. ACQUARONE}
\affiliation
{IMEM-CNR and Dipartimento di Fisica, Universit\`a di Parma, 43100 Parma, Italy\\E-mail: acquarone@fis.unipr.it}%

\author{C. I. VENTURA}
\affiliation{CONICET- Centro At\'omico Bariloche, 8400-Bariloche, Argentina\\
E-mail: ventura@cab.cnea.gov.ar}%



\begin{abstract}
{We focus on the definition of the unitary transformation leading to an effective second order Hamiltonian, inside 
degenerate eigensubspaces of the non-perturbed Hamiltonian. We shall prove, by working out in detail the
Su-Schrieffer-Heeger Hamiltonian case, that the presence of de\-ge\-ne\-ra\-te states, including fermions and bosons, 
which might seemingly pose an obstacle towards the determination of such "Fr\"{o}hlich-transformed" Hamiltonian, 
in fact does not:  we explicitly show how degenerate states may be harmlessly included in the treatment, as they 
contribute with va\-ni\-shing matrix elements to the effective Hamiltonian matrix. In such a way, one can use without 
difficulty the eigenvalues of the effective Hamiltonian to describe the renormalized energies of the real excitations 
in the interacting system. Our argument applies also to few-body systems where one may not invoke the thermodynamic 
limit to get rid of the "dangerous"  perturbation terms.}%
\end{abstract}

\maketitle

\vfill\eject

\section{Introduction.}

Effective Hamiltonians obtained by unitary transformations truncated to
second order in the perturbation, which here we shall label as of "Fr\"{o}%
hlich-type", are ubiquitous in condensed matter physics\cite{wagnerbook}.
Given a starting Hamiltonian $H=H_{0}+I$, with a \ basic part $H_{0}$ and a
perturbation $I$ , one may write the unitary transformed Hamiltonian
neglecting terms of third or higher order in the perturbation as: 
\begin{equation}
H^{{}}\equiv e^{\mathcal{R}}(H_{0}+I)e^{-\mathcal{R}}=H_{0}+I+\left[ 
\mathcal{R},H_{0}\right] +\left[ \mathcal{R},I\right] +\frac{1}{2}\left[ 
\mathcal{R},\left[ \mathcal{R},H_{0}\right] \right] +\mathcal{O}\left(
I^{3}\right)   \label{eq1}
\end{equation}

The antihermitian generator $\mathcal{R}$ \ of the unitary transformation $%
e^{\mathcal{R}}$\ is determined by imposing the vanishing of the first-order
terms 
\begin{equation}
I+\left[ \mathcal{R},H_{0}\right] =0  \label{eq3}
\end{equation}
so that the transformed second-order Hamiltonian $H_{{}}^{(2)}$ reads

\begin{equation}
H_{{}}^{(2)}=H_{0}+\frac{1}{2}\left[ \mathcal{R},I\right] +\mathcal{O}\left(
I^{3}\right)  \label{eq.Heffective}
\end{equation}

To solve Eq.\ref{eq3} for $\mathcal{R}$ , one may explicitate it on a
complete set of orthonormal eigenstates \ $|X_{\alpha}^{{}}\rangle$ of $%
H_{0} $ : 
\begin{equation}
H_{0}|X_{\alpha}^{{}}\rangle=\mathcal{E}_{\alpha}^{{}}|X_{\alpha}^{{}}%
\rangle\qquad\boldsymbol{1}=\sum_{\alpha}|X_{\alpha}^{{}}\rangle\langle
X_{\alpha}^{{}}|
\end{equation}
yielding the basic equation to determine the matrix elements of $\mathcal{R}$
\begin{equation}
\langle X_{\beta}|I|X_{\alpha}\rangle=\left( \mathcal{E}_{\beta}^{{}}-%
\mathcal{E}_{\alpha}^{{}}\right) \langle X_{\beta}|\mathcal{R}|X_{\alpha
}\rangle  \label{matel.Fgeneral}
\end{equation}

In solving Eq.\ref{matel.Fgeneral} one has to distinguish between the cases
of \ non-degenerate or degenerate eigenstates. For non-degenerate cases, the
matrix elements of $\mathcal{R}$ are directly obtained and the second-order
Hamiltonian is straightforwardly derived, as usually done. The \ presence of
degenerate eigenstates of $\ H_{0}$ \ requires a special analysis. Namely,
if $\mathcal{E}_{\alpha }^{{}}=\mathcal{E}_{\beta }^{{}}$ the right hand
side of the equation vanishes identically for any finite $\langle X_{\beta }|%
\mathcal{R}|X_{\alpha }\rangle ,$ and two possibilities appear, depending on
the perturbation $I$. If also $\langle X_{\beta }|I|X_{\alpha }\rangle =0$ ,
the left-hand side vanishes as well, \ and one gets no condition for the
corresponding matrix element of $\mathcal{R}$ between these degenerate
states. On the other hand, for $\langle X_{\beta }|I|X_{\alpha }\rangle \neq
0$, Eq.\ref{matel.Fgeneral} would make no sense except for the case of a
divergent $\langle X_{\beta }|\mathcal{R}|X_{\alpha }\rangle $. We will
focus on this problem, and explicitly show by working out in detail a
non-trivial example that a well defined unitary transformation can
nonetheless be defined when degenerate eigenstates are present (as\ first
observed in general by Wagner \cite{wagnerbook2}) , and through it the
second-order Hamiltonian be obtained. To fix ideas, we will consider the
case of a non-perturbed Hamiltonian $H_{0}$\ describing a system of
non-interacting bosons and fermions, in particular the Su-Schrieffer-Heeger
electron-phonon Hamiltonian.The degenerate states $|X_{\alpha }^{{}}\rangle $
and \ \ $|X_{\beta }^{{}}\rangle $ \ might differ in having one more, or one
less, boson, which is created or destroyed in a real (as opposed to virtual)
scattering process with the fermions, as described by perturbation $I$,
conserving the total energy between initial and final states. We show how
the generator $\mathcal{R}$ can be defined, such that the frequently used
procedure of identifying renormalized excitation energies of the real
interacting bosons with eigenvalues of their effective Hamiltonian, obtained
by averaging over the fermion wavefunction, can be applied without
difficulties.

\section{The Su-Schrieffer-Heeger electron-phonon Hamiltonian}

Let us work out in detail a specific example based on the
Su-Schrieffer-Heeger electron-phonon Hamiltonian \cite{wagnerbook}, for
which the non-interacting Hamiltonian reads: 
\begin{equation}
H_{0}^{{}}=\sum_{k\sigma}E_{k}^{{}}c_{k\sigma}^{\dagger}c_{k\sigma}^{{}}+%
\sum_{q}\hbar\Omega_{q}\left( b_{q}^{\dagger}b_{q}^{{}}+\frac{1}{2}\right)
\end{equation}
Introducing the number operators $n_{k\sigma}^{{}}=c_{k\sigma}^{\dagger
}c_{k\sigma}^{{}}$ , for itinerant electrons in states characterized by:
crystal momentum $k$ , spin $\sigma,$\ and energy $E_{k}^{{}},$ and $\nu
_{q}=b_{q}^{\dagger}b_{q}^{{}}$\ , for phonons in states characterized by
crystal momentum $q$ and energy $\Omega_{q}$\ , \ the generic eigenstate \ $%
|X_{\alpha}^{{}}\rangle=|X_{k,q,\sigma}^{{}}\rangle$\ of $H_{0}$ \ may be
written as a product of \ fermonic and bosonic orthonormal eigenstates: 
\begin{equation}
|X_{kq\sigma}^{{}}\rangle=|..n_{k\sigma}^{{}},n_{k+q,\sigma}^{{}},n_{k-q,%
\sigma}^{{}},\cdots\rangle.|..\nu_{q}^{{}},\nu_{-q}^{{}},\cdots\rangle 
\notag
\end{equation}
Above, we have explicitly written only the occupation numbers of the states
of interest for our discussion. For instance, the boson state indicated
simply as $|\nu_{q}..\rangle$, denotes the following $N$ -phonon state in
occupation number representation ($N$ different wavevectors): 
\begin{equation}
|\nu_{q}\rangle\Longleftrightarrow\left[ \frac{1}{\sqrt{\left\langle \nu
_{q}\right\rangle !}}\left( b_{q}^{\dagger}\right) ^{\nu_{q}}\right]
\prod_{p\neq q}\frac{1}{\sqrt{\left\langle \nu_{p}\right\rangle !}}\left(
b_{p}^{\dagger}\right) ^{\nu_{p}}|0\rangle
\end{equation}
while the orthonormal boson states $\left(
\langle\nu_{q}|\nu_{p}\rangle=\delta_{pq}\right) $ are such that: 
\begin{equation}
b_{q}^{{}}|\nu_{q}\rangle=\sqrt{\left\langle \nu_{q}\right\rangle }|\nu
_{q}-1\rangle\qquad b_{q}^{\dagger}|\nu_{q}\rangle=\sqrt{\left\langle \nu
_{q}\right\rangle +1}|\nu_{q}+1\rangle
\end{equation}

\bigskip

The exact eigenenergy $\mathcal{E}_{kq\sigma }^{{}}$\ \ of \ $|X_{kq\sigma
}^{{}}\rangle $ \ is: 
\begin{equation*}
\mathcal{E}_{kq\sigma }^{{}}=E_{k\sigma }^{{}}\left\langle n_{k\sigma
}^{{}}\right\rangle +E_{k+q,\sigma }^{{}}\left\langle n_{k+q,\sigma
}^{{}}\right\rangle +E_{k-q,\sigma }^{{}}\left\langle n_{k-q,\sigma
}^{{}}\right\rangle +\hbar \Omega _{q}\left\langle \nu _{q}^{{}}\right\rangle
\end{equation*}%
\begin{equation}
+\sum_{p\neq \left( k,k\pm q\right) }E_{p\sigma }^{{}}\left\langle
n_{p\sigma }^{{}}\right\rangle +\sum_{r\neq q}\hbar \Omega _{r}\left\langle
\nu _{r}^{{}}\right\rangle
\end{equation}

For this Hamiltonian, the perturbation term $I$ \ reads: 
\begin{equation}
I=\frac{1}{\sqrt{N}}\sum_{kq\sigma}\Gamma_{k,k-q}^{{}}c_{k\sigma}^{\dagger
}c_{k-q,\sigma}^{{}}\left( b_{-q}^{\dagger}+b_{q}^{{}}\right)  \label{def.Ia}
\end{equation}
where the bond-stretching interaction amplitude, resulting from a modulation
of the electron hopping $t_{lj}$ , for a lattice with a centre of inversion
is : 
\begin{equation}
\Gamma_{k,k-q}^{{}}=i\sum_{\langle lj\rangle}g_{lj}^{{}}\left\{ \sin\left[
\left( k-q\right) \Delta_{lj}^{{}}\right] -\sin\left[ k\Delta_{lj}^{{}}%
\right] \right\} \qquad g_{lj}^{{}}=\frac{\partial t_{lj}}{\partial\left(
R_{l}-R_{j}\right) }  \label{lin.3}
\end{equation}

Notice $\Gamma_{k,k-q}^{{}}=-\Gamma_{k,k-q}^{\ast}$ . Thus, rewriting $%
\sum_{kq\sigma}\Gamma_{k,k-q}^{{}}c_{k\sigma}^{\dagger}c_{k-q,%
\sigma}^{{}}b_{-q}^{\dagger}=\sum_{kq\sigma}\Gamma_{k,k-q}^{\ast}c_{k-q,%
\sigma}^{\dagger }c_{k,\sigma}^{{}}b_{q}^{\dagger}$ , we can decompose the
perturbation as

\begin{equation}
I=\frac{1}{\sqrt{N}}\sum_{kq\sigma}\left(
I_{kq\sigma}^{+}+I_{kq\sigma}^{-}\right)
\end{equation}

where 
\begin{equation}
I_{kq\sigma}^{+}=\Gamma_{k,k-q}^{\ast}c_{k-q,\sigma}^{\dagger}c_{k,%
\sigma}^{{}}b_{q}^{\dagger}\qquad\ \ \ \ \ \ \ \ \ I_{kq\sigma}^{-}=\Gamma
_{k,k-q}^{{}}c_{k\sigma}^{\dagger}c_{k-q,\sigma}^{{}}b_{q}^{{}}
\label{def.I kq}
\end{equation}

A given perturbation term $I_{kq,\sigma}^{\pm}$ yields non-vanishing results
only when applied to the specific state $|A_{kq\sigma}^{\pm}\rangle$ where: 
\begin{equation}
|A_{kq\sigma}^{+}\rangle=|1_{k\sigma}^{{}},n_{k+q,\sigma}^{{}},0_{k-q,\sigma
}^{{}}..\rangle|\nu_{q}^{{}},\nu_{-q}^{{}}..\rangle\qquad|A_{kq\sigma}^{-}%
\rangle=|0_{k\sigma}^{{}},1_{k-q,\sigma}^{{}},n_{k+q,\sigma}^{{}}..\rangle|%
\nu_{-q}^{{}},\nu_{q}^{{}}..\rangle  \label{def. | A+- kq>}
\end{equation}
Namely,%
\begin{equation}
I_{kq\sigma}^{+}|A_{kq\sigma}^{+}\rangle=\Gamma_{k,k-q}^{\ast}\sqrt {%
\left\langle \nu_{q}\right\rangle +1}|0_{k\sigma}^{{}},n_{k+q,%
\sigma}^{{}},1_{k-q,\sigma}^{{}}\rangle|\nu_{q}^{{}}+1,\nu_{-q}^{{}}\rangle%
\equiv \Gamma_{k,k-q}^{\ast}|B_{kq\sigma}^{+}\rangle  \label{I+A+}
\end{equation}%
\begin{equation}
I_{kq\sigma}^{-}|A_{kq\sigma}^{-}\rangle=\Gamma_{k,k-q}^{{}}\sqrt{%
\left\langle \nu_{q}\right\rangle }|1_{k\sigma}^{{}},n_{k+q,%
\sigma}^{{}},0_{k-q,\sigma}^{{}}\rangle|\nu_{q}^{{}}-1,\nu_{-q}^{{}}\rangle%
\equiv\Gamma_{k,k-q}^{{}}|B_{kq\sigma}^{-}\rangle  \label{I-A-}
\end{equation}
where we have introduced the notation $\ |B_{kq\sigma}^{\pm}\rangle$ for the
states resulting from applying the perturbation. Notice also that: $%
I_{k,q,\sigma}^{\pm}|A_{k,q,\sigma}^{\mp}\rangle=0$, and that states $%
|A_{kq\sigma}^{+}\rangle$ and \ $|B_{kq\sigma}^{+}\rangle$\ are orthogonal:%
\begin{equation}
\langle A_{kq\sigma}^{+}|B_{kq\sigma}^{+}\rangle=\frac{1}{%
\Gamma_{k,k-q}^{\ast}}\langle
A_{kq\sigma}^{+}|I_{kq\sigma}^{+}|A_{kq\sigma}^{+}\rangle=0
\end{equation}
To fix ideas, in the following we will solve in detail the problem for the
case of two degenerate eigenstates of the unperturbed Hamiltonian. The
argumentation can be straightforwardly extended for larger size of the
degenerate eigensubspace, without altering the conclusions.

\section{\protect\bigskip The case of two degenerate states.}

Let us now assume that the spectrum of non-interacting energies\ of the
system is such that, given the state $|A_{kq\sigma}^{+}\rangle=|1_{k%
\sigma}^{{}},n_{k+q,\sigma}^{{}},0_{k-q,\sigma}^{{}}\rangle|\nu_{q}^{{}},%
\nu_{-q}^{{}}\rangle$ with energy $\mathcal{E}_{kq\sigma}^{A}$, a phonon
exists such that $|A_{kq\sigma}^{+}\rangle$ is degenerate with the state $%
|B_{kq\sigma}^{+}\rangle=\sqrt{\left\langle \nu_{q}\right\rangle +1}%
|0_{k\sigma}^{{}},n_{k+q,\sigma}^{{}},1_{k-q,\sigma}^{{}}\rangle|%
\nu_{q}^{{}}+1,\nu_{-q}^{{}}\rangle$\ . Thus $\mathcal{E}_{kq\sigma}^{B}=%
\mathcal{E}_{kq\sigma}^{A}$, which\ \ implies $E_{k}^{{}}=E_{k-q}^{{}}+\hbar%
\Omega_{q}.$ \ In this case, we might be in trouble with Eq.\ref%
{matel.Fgeneral}, as mentioned in the Introduction. \ 

To tackle the problem, it is convenient to rewrite the condition for the
transformation generator $\mathcal{R}$ , Eq.\ref{eq3} , specifically
isolating in $I\ \ $the "dangerous" terms $I_{kq\sigma }^{+}+I_{kq\sigma
}^{-}$: 
\begin{equation}
I=\frac{1}{\sqrt{N}}\sum_{pr\tau }\left( I_{pr\tau }^{+}+I_{pr\tau
}^{-}\right) \left( 1-\delta _{pk}\delta _{rq}\delta _{\tau \sigma }\right) +%
\frac{1}{\sqrt{N}}\left( I_{kq\sigma }^{+}+I_{kq\sigma }^{-}\right) \equiv
I_{1}+\mathcal{I}_{kq\sigma }  \label{eq.18}
\end{equation}

Analogously, we write for the generator: $\mathcal{R}=\mathcal{R}_{1}+%
\mathcal{R}_{kq\sigma }$ , explicitating its restriction to the degenerate
subspace in last term. Notice that the \ commutators $\left[ I_{1},\mathcal{I%
}_{kq\sigma }\right] $, $\left[ \mathcal{R}_{1},\mathcal{R}_{kq\sigma }%
\right] $ , $\left[ \mathcal{R}_{1},\mathcal{I}_{kq\sigma }\right] $ and $%
\left[ I_{1},\mathcal{R}_{kq\sigma }\right] $ vanish, since the operators to
be commuted involve different wavevectors. Thus, the transformed Hamiltonian
of Eq.\ref{eq1} \ now reads: 
\begin{equation*}
H^{{}}\equiv e^{\left( \mathcal{R}_{1}+\mathcal{R}_{kq\sigma }\right)
}(H_{0}+I_{1}+\mathcal{I}_{kq\sigma })e^{-\left( \mathcal{R}_{1}+\mathcal{R}%
_{kq\sigma }\right) }=
\end{equation*}%
\begin{equation*}
=H_{0}+I_{1}+\mathcal{I}_{kq\sigma }+\left[ \left( \mathcal{R}_{1}+\mathcal{R%
}_{kq\sigma }\right) ,H_{0}\right] +\left[ \mathcal{R}_{1},I_{1}\right] +%
\left[ \mathcal{R}_{kq\sigma },\mathcal{I}_{kq\sigma }\right] 
\end{equation*}%
\begin{equation}
+\frac{1}{2}\left[ \mathcal{R}_{1},\left[ \mathcal{R}_{1},H_{0}\right] %
\right] +\frac{1}{2}\left[ \mathcal{R}_{1},\left[ \mathcal{R}_{kq\sigma
},H_{0}\right] \right] +\frac{1}{2}\left[ \mathcal{R}_{kq\sigma },\left[ 
\mathcal{R}_{1},H_{0}\right] \right] +\frac{1}{2}\left[ \mathcal{R}%
_{kq\sigma },\left[ \mathcal{R}_{kq\sigma },H_{0}\right] \right] +\mathcal{O}%
\left( I^{3}\right)   \label{eq20}
\end{equation}

The condition to be imposed in order\ to eliminate the terms linear in the
perturbation takes the form:%
\begin{equation}
I_{1}+\mathcal{I}_{kq\sigma}+\left[ \left( \mathcal{R}_{1}+\mathcal{R}%
_{kq\sigma}\right) ,H_{0}\right] =0
\end{equation}

The above equation for $\mathcal{R}$ can further be split into two
independent constraints%
\begin{equation}
I_{1}+\left[ \mathcal{R}_{1},H_{0}\right] =0\Longrightarrow\left[ \mathcal{R}%
_{1},H_{0}\right] =-I_{1}
\end{equation}%
\begin{equation}
\mathcal{I}_{kq\sigma}+\left[ \mathcal{R}_{kq\sigma},H_{0}\right]
=0\Longrightarrow\left[ \mathcal{R}_{kq\sigma},H_{0}\right] =-\mathcal{I}%
_{kq\sigma}
\end{equation}

Therefore, using these two constraints in Eq.\ref{eq20} and dropping all
vanishing commutators, it follows that to second order one has: 
\begin{equation}
H^{(2)}=H_{0}+\frac{1}{2}\left[ \mathcal{R}_{1},I_{1}\right] +\frac{1}{2}%
\left[ \mathcal{R}_{kq\sigma},\mathcal{I}_{kq\sigma}\right] +\mathcal{O}%
\left( I^{3}\right)  \label{eq.Heffective0}
\end{equation}
where the matrix elements of commutator $\left[ \mathcal{R}_{1},I_{1}\right] 
$ can be evaluated safely even over the pair of degenerate states. In the
following, we shall focus on the matrix elements of the potentially
"dangerous" commutator \ $\left[ \mathcal{R}_{kq\sigma},\mathcal{I}%
_{kq\sigma}\right] .$

\subsection{The action of the perturbation on the degenerate states.}

As $|B_{kq\sigma }^{+}\rangle $ is not normalized, let's introduce the
corresponding normalized eigenstate $|\mathcal{B}_{kq\sigma }^{+}\rangle $
as: 
\begin{equation}
|\mathcal{B}_{kq\sigma }^{+}\rangle =\frac{1}{\sqrt{\left\langle \nu
_{q}\right\rangle +1}}|B_{kq\sigma }^{+}\rangle =|0_{k\sigma
}^{{}},n_{k+q,\sigma }^{{}},1_{k-q,\sigma }^{{}}\rangle |\nu _{q}^{{}}+1,\nu
_{-q}^{{}}\rangle  \label{def.cal | B kq>}
\end{equation}

Let us evaluate the matrix elements of $\left( I_{kq\sigma}^{+}+I_{kq\sigma
}^{-}\right) $ inside the \ degenerate subspace spanned by: $|A_{kq\sigma
}^{+}\rangle,$\ $|\mathcal{B}_{kq\sigma}^{+}\rangle$ .The diagonal elements
vanish by orthogonality: $\langle A_{kq\sigma}^{+}|\left(
I_{kq\sigma}^{+}+I_{kq\sigma}^{-}\right) |A_{kq\sigma}^{+}\rangle=\langle%
\mathcal{B}_{kq\sigma}^{+}|\left( I_{kq\sigma}^{+}+I_{kq\sigma}^{-}\right) |%
\mathcal{B}_{kq\sigma}^{+}\rangle=0$ \ since the perturbation changes the
number of bosons, while:%
\begin{equation}
\langle\mathcal{B}_{kq\sigma}^{+}|\left(
I_{kq\sigma}^{+}+I_{kq\sigma}^{-}\right)
|A_{kq\sigma}^{+}\rangle=\Gamma_{k,k-q}^{\ast}\sqrt{\left\langle
\nu_{q}\right\rangle +1}=\left[ \langle A_{kq\sigma}^{+}|\left( I_{kq\sigma
}^{+}+I_{kq\sigma}^{-}\right) |\mathcal{B}_{kq\sigma}^{+}\rangle\right]
^{\ast}  \label{Iqaction}
\end{equation}

\subsection{The perturbation-split equivalent states and the corresponding
matrix elements.}

By a rotation inside the degenerate subspace, we now diagonalize the
restriction of the perturbation$\frac{1}{\sqrt{N}}($ $I_{kq\sigma}^{+}+I_{kq%
\sigma}^{-})$, there. From its eigenvalues, we obtain the respective
eigenenergies of the Hamiltonian including first-order perturbative
corrections: 
\begin{equation}
E_{kq\sigma}^{\pm}=\mathcal{E}_{kq\sigma}^{A}\pm\left\vert \langle \mathcal{B%
}_{kq\sigma}^{+}|\frac{1}{\sqrt{N}}\left(
I_{kq\sigma}^{+}+I_{kq\sigma}^{-}\right) |A_{kq\sigma}^{+}\rangle\right\vert
=\mathcal{E}_{kq\sigma}^{A}\pm\frac{1}{\sqrt{N}}\left\vert
\Gamma_{k,k-q}^{{}}\right\vert \sqrt{\left\langle \nu_{q}\right\rangle +1}
\end{equation}

\bigskip with respective orthonormal eigenvectors of $H_{0}:$ 
\begin{align}
|\psi _{kq\sigma }^{+}\rangle & =\frac{1}{\sqrt{2}}\left[ \lambda
|A_{kq\sigma }^{+}\rangle +|\mathcal{B}_{kq\sigma }^{+}\rangle \right]
\qquad |\psi _{kq\sigma }^{-}\rangle =\frac{1}{\sqrt{2}}\left[ \lambda
|A_{kq\sigma }^{+}\rangle -|\mathcal{B}_{kq\sigma }^{+}\rangle \right]
\qquad  \\
\quad \lambda & \equiv \frac{\Gamma _{k,k-q}^{{}}}{\left\vert \Gamma
_{k,k-q}^{{}}\right\vert }=i\lambda ^{\prime }\quad \mathrm{where}\quad
\lambda ^{\prime }\equiv sgn\left[ \sum_{\langle lj\rangle
}g_{lj}^{{}}\left\{ \sin \left[ \left( k-q\right) \Delta _{lj}^{{}}\right]
-\sin \left[ k\Delta _{lj}^{{}}\right] \right\} \right] 
\end{align}

The diagonal matrix elements of "dangerous" perturbation terms in the $%
|\psi_{kq\sigma}^{\pm}\rangle$ subspace are$:$ 
\begin{equation*}
\langle\psi_{kq\sigma}^{+}|\left( I_{kq\sigma}^{+}+I_{kq\sigma}^{-}\right)
|\psi_{kq\sigma}^{+}\rangle=-\left\vert \Gamma_{k,k-q}^{{}}\right\vert \sqrt{%
\left\langle \nu_{q}\right\rangle +1}
\end{equation*}%
\begin{equation}
\langle\psi_{kq\sigma}^{-}|\left( I_{kq\sigma}^{+}+I_{kq\sigma}^{-}\right)
|\psi_{kq\sigma}^{-}\rangle=\left\vert \Gamma_{k,k-q}^{{}}\right\vert \sqrt{%
\left\langle \nu_{q}\right\rangle +1}  \label{Pert.diag.matel}
\end{equation}

while the off-diagonal ones vanish.

\subsection{Matrix elements of the effective Hamiltonian in the $|\protect%
\psi _{kq\protect\sigma}^{+}\rangle$ $,$ $|\protect\psi_{kq\protect\sigma%
}^{-}\rangle$ subspace.}

Of interest is the effect that the "dangerous" perturbation term might have
on the matrix elements of $H^{(2)}$ of Eq.\ref{eq.Heffective0}. We will show
that, with an appropriate definition of $\mathcal{R}$, all matrix elements
of $\left[ \mathcal{R}_{kq\sigma }^{{}},\mathcal{I}_{kq\sigma }\right] $ in
the subspace spanned by $|\psi _{kq\sigma }^{+}\rangle $ and $|\psi
_{kq\sigma }^{-}\rangle $ vanish.

For the evaluation of the off-diagonal elements of $\left[ \mathcal{R}%
_{kq\sigma }^{{}},\mathcal{I}_{kq\sigma }\right] \quad $it will be
convenient to use the following decomposition of the identity
operator:\bigskip 
\begin{equation*}
\mathbf{1=}\sum_{pr\sigma }|X_{pr\sigma }^{{}}\rangle \langle X_{pr\sigma
}^{{}}|\left( 1-\langle A_{pr\sigma }^{+}|X_{pr\sigma }^{{}}\rangle \right)
\left( 1-\langle \mathcal{B}_{pr\sigma }^{+}|X_{pr\sigma }^{{}}\rangle
\right)
\end{equation*}%
\begin{equation}
+|\psi _{kq\sigma }^{+}\rangle \langle \psi _{kq\sigma }^{+}|+|\psi
_{kq\sigma }^{-}\rangle \langle \psi _{kq\sigma }^{-}|  \label{identity}
\end{equation}%
where use has been made of the equality: $|\psi _{kq\sigma }^{+}\rangle
\langle \psi _{kq\sigma }^{+}|+|\psi _{kq\sigma }^{-}\rangle \langle \psi
_{kq\sigma }^{-}|=|A_{kq\sigma }^{+}\rangle \langle A_{kq\sigma }^{+}|+|%
\mathcal{B}_{kq\sigma }^{+}\rangle \langle \mathcal{B}_{kq\sigma }^{+}|.$
For simplicity, in the following we will assume that except for the subspace
explicitated above, $\ H_{0}$ possesses no other degenerate eigenstates.

We have:%
\begin{equation*}
\langle \psi _{kq\sigma }^{+}|\left[ \mathcal{R}_{kq\sigma }^{{}},\mathcal{I}%
_{kq\sigma }\right] |\psi _{kq\sigma }^{-}\rangle =
\end{equation*}
\begin{equation}
=\frac{1}{\sqrt{N}}\left[ \langle \psi _{kq\sigma }^{+}|\mathcal{R}%
_{kq\sigma }^{{}}\left( I_{kq\sigma }^{+}+I_{kq\sigma }^{-}\right) |\psi
_{kq\sigma }^{-}\rangle -\langle \psi _{kq\sigma }^{+}|\left( I_{kq\sigma
}^{+}+I_{kq\sigma }^{-}\right) \mathcal{R}_{kq\sigma }^{{}}|\psi _{kq\sigma
}^{-}\rangle \right]  \label{HeffPsi+-}
\end{equation}

Inserting the identity operator $\mathbf{1}$, the first contribution to the
rhs of Eq.\ref{HeffPsi+-} , reads: 
\begin{equation*}
\langle \psi _{kq\sigma }^{+}|\mathcal{R}_{kq\sigma }^{{}}\left( \mathbf{1}%
\right) \left( I_{kq\sigma }^{+}+I_{kq\sigma }^{-}\right) |\psi _{kq\sigma
}^{-}\rangle =
\end{equation*}%
\begin{equation*}
=\sum_{pr\sigma }\left( 1-\langle A_{pr\sigma }^{+}|X_{pr\sigma
}^{{}}\rangle \right) \left( 1-\langle \mathcal{B}_{pr\sigma
}^{+}|X_{pr\sigma }^{{}}\rangle \right) \langle \psi _{kq\sigma }^{+}|%
\mathcal{R}_{kq\sigma }^{{}}|X_{pr\sigma }^{{}}\rangle \langle X_{pr\sigma
}^{{}}|\left( I_{kq\sigma }^{+}+I_{kq\sigma }^{-}\right) |\psi _{kq\sigma
}^{-}\rangle +
\end{equation*}%
\begin{equation}
+\langle \psi _{kq\sigma }^{+}|\mathcal{R}_{kq\sigma }^{{}}|\psi _{kq\sigma
}^{+}\rangle \langle \psi _{kq\sigma }^{+}|\left( I_{kq\sigma
}^{+}+I_{kq\sigma }^{-}\right) |\psi _{kq\sigma }^{-}\rangle +\langle \psi
_{kq\sigma }^{+}|\mathcal{R}_{kq\sigma }^{{}}|\psi _{kq\sigma }^{-}\rangle
\langle \psi _{kq\sigma }^{-}|\left( I_{kq\sigma }^{+}+I_{kq\sigma
}^{-}\right) |\psi _{kq\sigma }^{-}\rangle  \label{HeffPsi+(1)}
\end{equation}

For the first term, notice that $\sum_{pr\sigma }$\ involves only
eigenstates outside the degenerate subspace specified, while, using Eqs.\ref%
{I+A+} , \ref{def.cal | B kq>} and \ref{Iqaction}, one has:%
\begin{equation*}
\langle X_{pr\sigma }^{{}}|\left( I_{kq\sigma }^{+}+I_{kq\sigma }^{-}\right)
|\psi _{kq\sigma }^{-}\rangle =
\end{equation*}
\begin{equation}
=\frac{\lambda }{\sqrt{2}}\Gamma _{k,k-q}^{\ast }\sqrt{\left\langle \nu
_{q}\right\rangle +1}\langle X_{pr\sigma }^{{}}|\mathcal{B}_{kq\sigma
}^{+}\rangle -\frac{1}{\sqrt{2}}\Gamma _{k,k-q}^{{}}\sqrt{\left\langle \nu
_{q}\right\rangle +1}\langle X_{pr\sigma }^{{}}|A_{kq\sigma }^{+}\rangle
\equiv 0  \label{zeroSum}
\end{equation}%
where the last equality follows from orthogonality, in both scalar products.

The next term in Eq.\ref{HeffPsi+(1)} vanishes because \ $\langle \psi
_{kq\sigma }^{+}|\left( I_{kq\sigma }^{+}+I_{kq\sigma }^{-}\right) |\psi
_{kq\sigma }^{-}\rangle =0$\ . The last term in Eq.\ref{HeffPsi+(1)} , due
to Eq.\ref{Pert.diag.matel},reads:%
\begin{equation*}
\langle \psi _{kq\sigma }^{+}|\mathcal{R}_{kq\sigma }^{{}}|\psi _{kq\sigma
}^{-}\rangle \langle \psi _{kq\sigma }^{-}|\left( I_{kq\sigma
}^{+}+I_{kq\sigma }^{-}\right) |\psi _{kq\sigma }^{-}\rangle =
\end{equation*}%
\begin{equation}
=-\left\vert \Gamma _{k,k-q}^{{}}\right\vert \sqrt{\left\langle \nu
_{q}\right\rangle +1\quad }\langle \psi _{kq\sigma }^{+}|\mathcal{R}%
_{kq\sigma }^{{}}|\psi _{kq\sigma }^{-}\rangle
\end{equation}

The off-diagonal matrix elements of $\mathcal{R}_{kq\sigma}^{{}}$ are
required to satisfy the condition given by Eq.\ref{matel.Fgeneral}. But due
to the diagonalization of the perturbation performed, and $%
H_{0}.|\psi_{kq\sigma }^{\pm}\rangle=\mathcal{E}_{kq\sigma}^{A}|\psi_{kq%
\sigma}^{\pm}\rangle$ , one is in the case in which of both sides of Eq.\ref%
{matel.Fgeneral} are identically zero. Thus, one in fact has the freedom to
choose the value of $\ \langle\psi_{kq\sigma}^{+}|\mathcal{R}%
_{kq\sigma}^{{}}|\psi_{kq\sigma}^{-}\rangle$ \bigskip: \ as zero, \ in
particular. In the following, we will justify this choice as a \ reasonable
and most convenient one.

\bigskip First, one can justify that choice by an analytical continuity
argument. \ Having performed the diagonalization of the perturbation, $%
\langle \psi _{kq\sigma }^{+}|\left( I_{kq\sigma }^{+}+I_{kq\sigma
}^{-}\right) |\psi _{kq\sigma }^{-}\rangle =0$ \ holds exactly for any value
of $\left\vert \Gamma _{k,k-q}^{{}}\right\vert \geq 0$. If \ in Eq.\ref%
{matel.Fgeneral} one would replace the zero-order (degenerate, here)
eigenenergies difference \ by the vanishing perturbation limit \ of the
first-order eigenenergies splitting, 
\begin{equation*}
0=\lim_{\left\vert \Gamma \right\vert \rightarrow 0}\left( E_{kq\sigma
}^{+}-E_{kq\sigma }^{-}\right) \langle \psi _{kq\sigma }^{+}|\mathcal{R}%
_{kq\sigma }^{{}}|\psi _{kq\sigma }^{-}\rangle =
\end{equation*}%
\begin{equation}
=\lim_{\left\vert \Gamma \right\vert \rightarrow 0}\frac{1}{\sqrt{N}}%
\left\vert \Gamma _{k,k-q}^{{}}\right\vert \sqrt{\left\langle \nu
_{q}\right\rangle +1\ }\langle \psi _{kq\sigma }^{+}|\mathcal{R}_{kq\sigma
}^{{}}|\psi _{kq\sigma }^{-}\rangle \quad \left( \left\vert \Gamma
_{k,k-q}^{{}}\right\vert \geq 0\right)
\end{equation}%
one would see that in the whole neighbourhood of $\left\vert \Gamma
_{k,k-q}^{{}}\right\vert =0\quad $the consistent definition of the
transformation would be:%
\begin{equation}
\langle \psi _{kq\sigma }^{+}|\mathcal{R}_{kq\sigma }^{{}}|\psi _{kq\sigma
}^{-}\rangle =0\qquad \left( \left\vert \Gamma _{k,k-q}^{{}}\right\vert
>0\right)  \label{choice}
\end{equation}%
\bigskip By continuity, it thus seems reasonable \ to choose $\langle \psi
_{kq\sigma }^{+}|\mathcal{R}_{kq\sigma }^{{}}|\psi _{kq\sigma }^{-}\rangle
=0,$ \ \ also for $\left\vert \Gamma _{k,k-q}^{{}}\right\vert =0.$

Further, as a hand-waving argument, one could also mention that since $%
\mathcal{R}_{kq\sigma}^{{}}$ has the same operatorial structure as $\left(
I_{kq\sigma}^{+}+I_{kq\sigma}^{-}\right) $, one might expect the two objects
to have similarly vanishing off-diagonal matrix elements in the $|\psi
_{kq\sigma}^{\pm}\rangle$ subspace.

Going back to Eq.\ref{HeffPsi+-}, with our choice we thus have a vanishing
contribution for the first term of the commutator. To evaluate the second
term of the commutator we again introduce the identity decomposition
previously used, yielding:%
\begin{equation*}
\langle \psi _{kq\sigma }^{+}|\left( I_{kq\sigma }^{+}+I_{kq\sigma
}^{-}\right) \left( \mathbf{1}\right) \mathcal{R}_{kq\sigma }^{{}}|\psi
_{kq\sigma }^{-}\rangle =
\end{equation*}%
\begin{equation*}
=\sum_{pr\sigma }\left( 1-\langle A_{pr\sigma }^{+}|X_{pr\sigma
}^{{}}\rangle \right) \left( 1-\langle \mathcal{B}_{pr\sigma
}^{+}|X_{pr\sigma }^{{}}\rangle \right) \langle \psi _{kq\sigma }^{+}|\left(
I_{kq\sigma }^{+}+I_{kq\sigma }^{-}\right) |X_{pr\sigma }^{{}}\rangle
\langle X_{pr\sigma }^{{}}|\mathcal{R}_{kq\sigma }^{{}}|\psi _{kq\sigma
}^{-}\rangle +
\end{equation*}%
\begin{equation}
+\langle \psi _{kq\sigma }^{+}|\left( I_{kq\sigma }^{+}+I_{kq\sigma
}^{-}\right) |\psi _{kq\sigma }^{+}\rangle \langle \psi _{kq\sigma }^{+}|%
\mathcal{R}_{kq\sigma }^{{}}|\psi _{kq\sigma }^{-}\rangle +\langle \psi
_{kq\sigma }^{+}|\left( I_{kq\sigma }^{+}+I_{kq\sigma }^{-}\right) |\psi
_{kq\sigma }^{-}\rangle \langle \psi _{kq\sigma }^{-}|\mathcal{R}_{kq\sigma
}^{{}}|\psi _{kq\sigma }^{-}\rangle  \label{eq.72}
\end{equation}

\bigskip Here, the first term is zero by orthogonality, analogously to Eq.%
\ref{zeroSum}, and having chosen to define: $\langle \psi _{kq\sigma }^{+}|%
\mathcal{R}_{kq\sigma }^{{}}|\psi _{kq\sigma }^{-}\rangle =0,$ the second
term of Eq.\ref{eq.72} vanishes as well. Finally, the last term also
vanishes, because $|\psi _{kq\sigma }^{\pm }\rangle $ are orthogonal
eigenvectors of the perturbation.

Therefore, Eq.\ref{HeffPsi+-} yields \ \textit{vanishing off-diagonal matrix
elements }of $\left[ \mathcal{R}_{kq\sigma }^{{}},\mathcal{I}_{kq\sigma }%
\right] .$

Let us now evaluate its \textit{diagonal elements}. We have to consider%
\begin{equation*}
\langle \psi _{kq\sigma }^{+}|\ \left[ \mathcal{R}_{kq\sigma }^{{}},\mathcal{%
I}_{kq\sigma }\right] |\psi _{kq\sigma }^{+}\rangle =
\end{equation*}
\begin{equation}
=\frac{1}{\sqrt{N}}\left[ \langle \psi _{kq\sigma }^{+}|\mathcal{R}%
_{kq\sigma }^{{}}\left( I_{kq\sigma }^{+}+I_{kq\sigma }^{-}\right) |\psi
_{kq\sigma }^{+}\rangle -\langle \psi _{kq\sigma }^{+}|\left( I_{kq\sigma
}^{+}+I_{kq\sigma }^{-}\right) \mathcal{R}_{kq\sigma }^{{}}|\psi _{kq\sigma
}^{+}\rangle \right]  \label{diag}
\end{equation}

Again, inserting the identity decomposition, the first contribution to the
commutator in Eq.\ref{diag} \ reads: 
\begin{equation*}
\langle \psi _{kq\sigma }^{+}|\mathcal{R}_{kq\sigma }^{{}}\left( I_{kq\sigma
}^{+}+I_{kq\sigma }^{-}\right) |\psi _{kq\sigma }^{+}\rangle =\langle \psi
_{kq\sigma }^{+}|\mathcal{R}_{kq\sigma }^{{}}\left( \mathbf{1}\right) \left(
I_{kq\sigma }^{+}+I_{kq\sigma }^{-}\right) |\psi _{kq\sigma }^{+}\rangle =
\end{equation*}%
\begin{equation*}
=\sum_{pr\sigma }\left( 1-\langle A_{pr\sigma }^{{}}|X_{pr\sigma
}^{{}}\rangle \right) \left( 1-\langle \mathcal{B}_{pr\sigma
}^{{}}|X_{pr\sigma }^{{}}\rangle \right) \langle \psi _{kq\sigma }^{+}|%
\mathcal{R}_{kq\sigma }^{{}}|X_{pr\sigma }^{{}}\rangle \langle X_{pr\sigma
}^{{}}|\left( I_{kq\sigma }^{+}+I_{kq\sigma }^{-}\right) |\psi _{kq\sigma
}^{+}\rangle +
\end{equation*}%
\begin{equation*}
+\langle \psi _{kq\sigma }^{+}|\mathcal{R}_{kq\sigma }^{{}}|\psi _{kq\sigma
}^{+}\rangle \langle \psi _{kq\sigma }^{+}|\left( I_{kq\sigma
}^{+}+I_{kq\sigma }^{-}\right) |\psi _{kq\sigma }^{+}\rangle
\end{equation*}%
\begin{equation}
+\langle \psi _{kq\sigma }^{+}|\mathcal{R}_{kq\sigma }^{{}}|\psi _{kq\sigma
}^{-}\rangle \langle \psi _{kq\sigma }^{-}|\left( I_{kq\sigma
}^{+}+I_{kq\sigma }^{-}\right) |\psi _{kq\sigma }^{+}\rangle \equiv 0
\end{equation}%
The last equality follows since in the first term\ the matrix elements $%
\langle X_{pr\sigma }^{{}}|\left( I_{kq\sigma }^{+}+I_{kq\sigma }^{-}\right)
|\psi _{kq\sigma }^{+}\rangle $ vanish by orthogonality condition, as we
have shown when evaluating the off-diagonal matrix elements; in the second
term the diagonal element $\langle \psi _{kq\sigma }^{+}|\mathcal{R}%
_{kq\sigma }^{{}}|\psi _{kq\sigma }^{+}\rangle =0$ , due to antihermiticity
of $\mathcal{R}_{kq\sigma }^{{}}$\ ; \ and, in the last contribution, the
perturbation diagonalization yields $\langle \psi _{kq\sigma }^{+}|\left(
I_{kq\sigma }^{+}+I_{kq\sigma }^{-}\right) |\psi _{kq\sigma }^{-}\rangle $ $%
=0.$

Analogously, it is immediate to demonstrate that the second term of the
commutator also vanishes, leading to:%
\begin{equation}
\langle\psi_{kq\sigma}^{+}|\left[ \mathcal{R}_{kq\sigma}^{{}},\left(
I_{kq\sigma}^{+}+I_{kq\sigma}^{-}\right) \right] |\psi_{kq\sigma}^{+}%
\rangle=0
\end{equation}

Using the same arguments, one can easily show that also the other diagonal
element vanishes: $\ \langle\psi_{kq\sigma}^{-}|\left[ \mathcal{R}_{kq\sigma
}^{{}},\left( I_{kq\sigma}^{+}+I_{kq\sigma}^{-}\right) \right]
|\psi_{kq\sigma}^{-}\rangle$ $=0$.

In conclusion, all contributions of the "dangerous " perturbation terms to
the second-order transformed Hamiltonian have vanishing matrix elements \
inside the degenerate subspace, if one avails oneself of the freedom one has
to complete the definition of generator $\mathcal{R}$ and chooses that its
restriction to the degenerate subspace is diagonal in the same basis of
eigenvectors of the perturbation, there. This conclusion is clearly not
dependent on the size of the degenerate eigensubspace. \ In our example, the
choice is: $\ \langle\psi_{kq\sigma}^{+}|\mathcal{R}_{kq\sigma}^{{}}|%
\psi_{kq\sigma}^{-}\rangle=0,$ where $|\psi_{kq\sigma}^{\pm}\rangle$\
diagonalize the perturbation restriction to the two-dimensional degenerate
subspace. Therefore, the eigenvalues of the complete Hamiltonian in the
degenerate subspace only have contributions from the "innocuous" commutator $%
\left[ R_{1},I_{1}\right] $. Using Eq.\ref{eq.Heffective0} , we can indeed
write: 
\begin{equation*}
\langle\psi_{kq\sigma}^{\alpha}|H^{(2)}|\psi_{kq\sigma}^{\beta}\rangle
=\langle\psi_{kq\sigma}^{\alpha}|\left\{ H_{0}+\frac{1}{2}\left[ R_{1},I_{1}%
\right] +\frac{1}{2\sqrt{N}}\left[ \mathcal{R}_{kq\sigma}^{{}},\left(
I_{kq\sigma}^{+}+I_{kq\sigma}^{-}\right) \right] \right\} |\psi_{kq\sigma
}^{\beta}\rangle=
\end{equation*}%
\begin{equation}
=\langle\psi_{kq\sigma}^{\alpha}|\left\{ H_{0}+\frac{1}{2}\left[ R_{1},I_{1}%
\right] \right\} |\psi_{kq\sigma}^{\beta}\rangle=\mathcal{E}%
^{A}\delta_{\alpha\beta}+\frac{1}{2}\langle\psi_{kq\sigma}^{\alpha}|\left[
R_{1},I_{1}\right] |\psi_{kq\sigma}^{\beta}\rangle\qquad\left( \alpha
,\beta=\pm\right)
\end{equation}

\section{Diagonalization of $H_{{}}^{(2)}$ in the\ degenerate $|\protect\psi %
_{kq\protect\sigma}^{\pm}\rangle$ subspace.}

At this stage, to use the "Fr\"{o}hlich"-transformed Hamiltonian in the
presence of degenerate states of $H_{0}$\ it would seem necessary to
actually determine the eigenvectors which diagonalize the perturbation
inside each degenerate subspace. We will now show that this is not the case:
in fact, that \ one can work with the transformed Hamiltonian only knowing
an eigenvector basis of the unperturbed Hamiltonian $H_{0}$, as is usually
the case.

To show this, we will explicitly diagonalize $H^{(2)}$ in the degenerate
subspace, spanned by $|\psi _{kq\sigma }^{\pm }\rangle $ in our example. Its
matrix elements are:%
\begin{equation*}
\langle \psi _{kq\sigma }^{+}|H^{(2)}{}^{{}}|\psi _{kq\sigma }^{+}\rangle =
\end{equation*}%
\begin{equation}
=\mathcal{E}_{kq\sigma }^{A}+\frac{1}{4}\left\{ \langle A_{kq\sigma }^{+}|%
\left[ R_{1},I_{1}\right] |A_{kq\sigma }^{+}\rangle +\langle \mathcal{B}%
_{kq\sigma }^{+}|\left[ R_{1},I_{1}\right] |\mathcal{B}_{kq\sigma
}^{+}\rangle \right\} -\frac{\lambda^{\prime }}{2} Im\langle
A_{kq\sigma }^{+}|\left[ R_{1},I_{1}\right] |\mathcal{B}_{kq\sigma
}^{+}\rangle
\end{equation}

\begin{equation*}
\langle \psi _{kq\sigma }^{-}|H^{(2)}|\psi _{kq\sigma }^{-}\rangle =
\end{equation*}%
\begin{equation}
=\mathcal{E}_{kq\sigma }^{A}+\frac{1}{4}\left\{ \langle A_{kq\sigma }^{+}|%
\left[ R_{1},I_{1}\right] |A_{kq\sigma }^{+}\rangle +\langle \mathcal{B}%
_{kq\sigma }^{+}|\left[ R_{1},I_{1}\right] |\mathcal{B}_{kq\sigma
}^{+}\rangle \right\} +\frac{\lambda^{\prime }}{2} Im\langle
A_{kq\sigma }^{+}|\left[ R_{1},I_{1}\right] |\mathcal{B}_{kq\sigma
}^{+}\rangle
\end{equation}

\begin{equation*}
\langle \psi _{kq\sigma }^{+}|H^{(2)}|\psi _{kq\sigma }^{-}\rangle =
\end{equation*}%
\begin{equation}
=\frac{1}{4}\left[ \langle A_{kq\sigma }^{+}|\left[ R_{1},I_{1}\right]
|A_{kq\sigma }^{+}\rangle -\langle \mathcal{B}_{kq\sigma }^{+}|\left[
R_{1},I_{1}\right] |\mathcal{B}_{kq\sigma }^{+}\rangle \right] +\frac{%
i\lambda^{\prime }}{2} Re\langle A_{kq\sigma }^{+}|\left[ R_{1},I_{1}%
\right] |\mathcal{B}_{kq\sigma }^{+}\rangle
\end{equation}

\begin{equation*}
\langle \psi _{kq\sigma }^{-}|H^{(2)}|\psi _{kq\sigma }^{+}\rangle =
\end{equation*}%
\begin{equation}
=\frac{1}{4}\left[ \langle A_{kq\sigma }^{+}|\left[ R_{1},I_{1}\right]
|A_{kq\sigma }^{+}\rangle -\langle \mathcal{B}_{kq\sigma }^{+}|\left[
R_{1},I_{1}\right] |\mathcal{B}_{kq\sigma }^{+}\rangle \right] -\frac{%
i\lambda^{\prime }}{2} Re\langle A_{kq\sigma }^{+}|\left[ R_{1},I_{1}%
\right] |\mathcal{B}_{kq\sigma }^{+}\rangle
\end{equation}

Now, the terms in $\left[ R_{1},I_{1}\right] $ can not transform $%
|A_{kq\sigma }^{+}\rangle $ into $|\mathcal{B}_{kq\sigma }^{+}\rangle $
because neither $R_{1}$ nor $I_{1}$ change the number of $|q\rangle $
phonons, this being instead the effect of \ $\left[ \mathcal{R}_{kq\sigma
}^{{}},\left( I_{kq\sigma }^{+}+I_{kq\sigma }^{-}\right) \right] $.
Therefore, off-diagonal matrix elements $\langle A_{kq\sigma }^{+}|\left[
R_{1},I_{1}\right] |\mathcal{B}_{kq\sigma }^{+}\rangle $ vanish. Conversely,
the diagonal elements $\langle A_{kq\sigma }^{+}|\left[ R_{1},I_{1}\right]
|A_{kq\sigma }^{+}\rangle ,\quad \langle \mathcal{B}_{kq\sigma }^{+}|\left[
R_{1},I_{1}\right] |\mathcal{B}_{kq\sigma }^{+}\rangle $ do not vanish \
because $\left[ R_{1},I_{1}\right] $ \ includes diagonal terms like $%
b_{p}^{\dagger }b_{p}^{{}}\left( 1-n_{r-p,\sigma }^{{}}\right) n_{r\sigma
}^{{}}\left( 1-\delta _{pq}\delta _{rk}\right) $ which do not necessarily
vanish in the $|A_{kq\sigma }^{+}\rangle ,\quad |\mathcal{B}_{kq\sigma
}^{+}\rangle $ states. \ It follows that:%
\begin{equation*}
\langle \psi _{kq\sigma }^{+}|H^{(2)}|\psi _{kq\sigma }^{+}\rangle =\langle
\psi _{kq\sigma }^{-}|H^{(2)}|\psi _{kq\sigma }^{-}\rangle =
\end{equation*}

\begin{equation}
=\mathcal{E}_{kq\sigma }^{A}+\frac{1}{4}\left\{ \langle A_{kq\sigma }^{+}|%
\left[ R_{1},I_{1}\right] |A_{kq\sigma }^{+}\rangle +\langle \mathcal{B}%
_{kq\sigma }^{+}|\left[ R_{1},I_{1}\right] |\mathcal{B}_{kq\sigma
}^{+}\rangle \right\} \rangle
\end{equation}

\begin{equation*}
\langle \psi _{kq\sigma }^{+}|H^{(2)}|\psi _{kq\sigma }^{-}\rangle =\langle
\psi _{kq\sigma }^{-}|H^{(2)}|\psi _{kq\sigma }^{+}\rangle =
\end{equation*}%
\begin{equation}
=\frac{1}{4}\left[ \langle A_{kq\sigma }^{+}|\left[ R_{1},I_{1}\right]
|A_{kq\sigma }^{+}\rangle -\langle \mathcal{B}_{kq\sigma }^{+}|\left[
R_{1},I_{1}\right] |\mathcal{B}_{kq\sigma }^{+}\rangle \right]
\end{equation}%
Thus, diagonalization of $H^{(2)}$in the degenerate subspace leads to the
following secular equation: 
\begin{equation}
\det 
\begin{bmatrix}
\left[ \mathcal{E}_{kq\sigma }^{A}+\left( A+B\right) -\lambda \right] & 
\left( A-B\right) \\ 
\left( A-B\right) & \left[ \mathcal{E}_{kq\sigma }^{A}+\left( A+B\right)
-\lambda \right]%
\end{bmatrix}%
=0
\end{equation}%
\bigskip where%
\begin{equation}
A\equiv \frac{1}{4}\langle A_{kq\sigma }^{+}|\left[ R_{1},I_{1}\right]
|A_{kq\sigma }^{+}\rangle \quad \quad ,\quad \quad B\equiv \frac{1}{4}%
\langle \mathcal{B}_{kq\sigma }^{+}|\left[ R_{1},I_{1}\right] |\mathcal{B}%
_{kq\sigma }^{+}\rangle
\end{equation}%
Therefore, the eigenvalues are:%
\begin{equation*}
\eta _{A}=\mathcal{E}_{kq\sigma }^{A}+2A=\mathcal{E}_{kq\sigma }^{A}+\frac{1%
}{2}\langle A_{kq\sigma }^{+}|\left[ R_{1},I_{1}\right] |A_{kq\sigma
}^{+}\rangle
\end{equation*}%
\begin{equation}
\eta _{B}=\mathcal{E}_{kq\sigma }^{A}+2B=\mathcal{E}_{kq\sigma }^{A}+\frac{1%
}{2}\langle \mathcal{B}_{kq\sigma }^{+}|\left[ R_{1},I_{1}\right] |\mathcal{B%
}_{kq\sigma }^{+}\rangle
\end{equation}

The eigenstate $|\Phi_{kq\sigma}^{\pm}\rangle=a|\psi_{kq\sigma}^{+}%
\rangle+b|\psi_{kq\sigma}^{-}\rangle$ has coefficients set by 
\begin{equation}
\left[ \mathcal{E}_{kq\sigma}^{A}+\left( A+B\right) -\eta_{A,B}\right]
a+\left( A-B\right) b=0
\end{equation}

or, $a=\pm b$ yielding $a=1/\sqrt{2}$ corresponding to the normalized
eigenvectors 
\begin{equation}
|\Phi_{kq\sigma}^{+}\rangle=\frac{1}{\sqrt{2}}\left[ |\psi_{kq\sigma}^{+}%
\rangle+|\psi_{kq\sigma}^{-}\rangle\right] =|A_{kq\sigma}^{+}\rangle\qquad%
\langle\Phi_{kq\sigma}^{+}|=\langle A_{kq\sigma}^{+}|
\end{equation}%
\begin{equation}
|\Phi_{kq\sigma}^{+}\rangle=\frac{1}{\sqrt{2}}\left[ |\psi_{kq\sigma}^{+}%
\rangle-|\psi_{kq\sigma}^{-}\rangle\right] =i|\mathcal{B}_{kq\sigma}^{+}%
\rangle\qquad\langle\Phi_{kq\sigma}^{+}|=-i\langle\mathcal{B}_{kq\sigma
}^{+}|
\end{equation}

Our final result is:%
\begin{equation}
\langle\Phi_{kq\sigma}^{+}|H_{eff}^{{}}|\Phi_{kq\sigma}^{+}\rangle=\langle
A_{kq\sigma}^{+}|H_{eff}^{{}}|A_{kq\sigma}^{+}\rangle=\mathcal{E}_{kq\sigma
}^{A}+\frac{1}{2}\langle A_{kq\sigma}^{+}|\left[ R_{1},I_{1}\right]
|A_{kq\sigma}^{+}\rangle
\end{equation}%
\begin{equation}
\langle\Phi_{kq\sigma}^{-}|H_{eff}^{{}}|\Phi_{kq\sigma}^{-}\rangle =\langle%
\mathcal{B}_{kq\sigma}^{+}|H_{eff}^{{}}|\mathcal{B}_{kq\sigma}^{+}\rangle=%
\mathcal{E}_{kq\sigma}^{A}+\frac{1}{2}\langle\mathcal{B}_{kq\sigma}^{+}|%
\left[ R_{1},I_{1}\right] |\mathcal{B}_{kq\sigma}^{+}\rangle
\end{equation}

The conclusion is that , in evaluating the spectrum of $H^{(2)}$one can use
the eigenstates of $H_{0}$ and simply ignore the "dangerous terms" $\left[ 
\mathcal{R}_{kq\sigma}^{{}},\left( I_{kq\sigma}^{+}+I_{kq\sigma}^{-}\right) %
\right] $, because only the "innocuous terms" \ $\left[ R_{1},I_{1}\right] $%
\ \ \ will modify the non-interacting spectrum.

\section{Conclusions.}

We have explicitly shown by working out in detail an example based on the
Su-Schrieffer-Heeger Hamiltonian, how to complete the definition of the
transformation generator when degenerate eigenstates of the non-perturbed
Hamiltonian are present. Namely, these states, which would apparently cause
a divergence in the effective pairing between fermions and bosons generated
by transforming "\`{a} la Fr\"{o}hlich" an interacting Hamiltonian, actually
contribute vanishing matrix elements to the effective Hamiltonian matrix.
The matrix elements of the latter can therefore be evaluated by using the
eigenstates of the basic $H_{0},$ and ignoring the "dangerous terms". The
conclusions do not depend on the size of the degenerate eigensubspace, as
our argumentation can be straightforwardly generalized for the case of more
than two degenerate states. Notice that, even if we have explicitated the
eigenstates $|X_{kq\sigma }^{+}\rangle $ as product of separate fermionic
and bosonic eigenfunctions, actually our reasoning holds for any structure
of $|X_{kq\sigma }^{+}\rangle .$ For non-factorized eigenstates, the
numerical coefficients in the formulae above would be different, but the \
conclusion would still be the same. The procedure has been successfully
applied also to other complex fermion-boson Hamiltonians, e.g. the one for
interacting ferromagnetic spin waves and electrons considered in \cite%
{MAandCV}. \ Thus, at low temperatures renormalized excitation energies of
the real interacting bosons can be described by the eigenvalues of the
effective second-order boson Hamiltonian, obtained by averaging over the
fermion ground state wavefunction.

As a final observation, notice that when we isolate the dangerous term in
the perturbation (Eq.\ref{eq.18}) it comes weighed by a factor $1/\sqrt{N}$.
So in the infinite-lattice limit this part might be assumed to give in any
case an irrelevant contribution . In recent times, a great deal of work has
been done on finite or low-dimensional systems (mesoscopic, quantum dots
etc.) where such a justification for neglecting the dangerous terms would not
hold. Our demonstration, however, holds even for a single-particle system,
because it concerns the matrix elements, irrespective of the kind of system
where they are evaluated.

\section*{Acknowledgments}

M. A. acknowledges stimulating criticism from M. A. Gusm\~{a}o (UFRGS; Porto
Alegre, Brasil), useful discussions with C. Alabiso (University of Parma,
Italy), and financial support from MIUR of Italy through PRIN\
2004022024\_005 "Disorder effects on the magnetic, fluctuative and
dissipative properties, and on the fluxon lattice in two-band
superconductors: experiments and theories". C.I.Ventura is a member of
Carrera del Investigador Cient\'{\i}fico of CONICET, and wishes to
acknowledge support from Consiglio Nazionale delle Ricerche (Short-Term
Mobility Grant). C.I.V. also thanks the Department of Physics of the
University of Parma, for its hospitality and support.


\begin{thebibliography}{9}
\bibitem{wagnerbook} M. Wagner:\ "\emph{Unitary Transformations in Solid
State Physics}" (North Holland, Amsterdam 1986), and references therein.

\bibitem{wagnerbook2} in pages 17-18 of Ref.\cite{wagnerbook}

\bibitem{MAandCV} C. I. Ventura and M. Acquarone, \emph{Phys. Rev.} B\textbf{%
70}, 184409-1/7 (2004).
\end{thebibliography}
\end{document}